# Measuring the Knowledge Base of an Economy in terms of Triple-Helix Relations among 'Technology, Organization, and Territory'

*Research Policy* (forthcoming)

Loet Leydesdorff,[1] Wilfred Dolfsma,[2] & Gerben van der Panne[3]

**Abstract**

Can the knowledge base of an economy be measured? In this study, we combine the perspective of regional economics on the interrelationships among technology, organization, and territory with the triple-helix model, and offer the mutual information in three dimensions as an indicator of the configuration. When this probabilistic entropy is negative, the configuration reduces the uncertainty that prevails at the systems level. Data about more than a million Dutch companies are used for testing the indicator. The data contain postal codes (geography), sector codes (proxy for technology), and firm sizes in terms of number of employees (proxy for organization). The configurations are mapped at three levels: national (NUTS-1), provincial (NUTS-2), and regional (NUTS-3). The levels are cross-tabled with the knowledge-intensive sectors and services. The results suggest that medium-tech sectors contribute to the knowledge base of an economy more than high-tech ones. Knowledge-intensive services have an uncoupling effect, but less so at the high-tech end of these services.

---

[1] *corresponding author*; Amsterdam School of Communications Research (ASCoR) University of Amsterdam, Kloveniersburgwal 48, NL-1012 CX Amsterdam, Tel.: +31-20- 525 6598; fax: +31-20- 525 3681,  loet@leydesdorff.net; http://www.leydesdorff.net .
[2] Erasmus University Rotterdam – FBK, PO Box 1738, NL-3000 DR  Rotterdam, w.dolfsma@fbk.eur.nl & MERIT - Maastricht University.
[3] Delft University of Technology, Economics of Innovation, Jaffalaan 5, 2628 BX  Delft, the Netherlands, g.vanderpanne@tbm.tudelft.nl

# 1. Introduction

Ever since evolutionary economists introduced the concept of a 'knowledge-based economy' (Foray & Lundvall, 1996; Abramowitz & David, 1996), the question of the measurement of this new type of economic coordination has come to the fore (Carter, 1996; OECD, 1996). Recently, Godin (forthcoming) argued that the concept itself has remained a rhetorical device because the development of specific indicators has failed. However, the concept of a 'knowledge-based economy' has been attractive to policy-makers at the level of the European Union, perhaps as an alternative to the '*national* systems of innovation' approach. For example, the European Summit of March 2000 in Lisbon was specifically held in order 'to strengthen employment, economic reform, and social cohesion in the transition to a knowledge-based economy' (European Commission, 2000; cf. European Commission, 2005).

Can something as elusive as the knowledge base of a system be measured? (Foray, 2004; Leydesdorff, 2001; Skolnikoff, 1993). Is a structural transformation of the economy at the global level indicated with potentially different effects in various world regions and nations? When originally proposing their program of studies at the OECD, David & Foray (1995, at p. 14) argued that the focus on *national* systems of innovation had placed too much emphasis on the organization of institutions and economic growth, and not enough on the distribution of knowledge itself. However, Lundvall's (1988) argument for considering the nation as a first candidate for defining innovation systems was carefully formulated in terms of an heuristics:

> The interdependency between production and innovation goes both ways. […] This interdependency between production and innovation makes it legitimate to take the national system of production as a starting point when defining a system of innovation. (Lundvall, 1988: 362)

The choice of the nation as a frame of reference enables the analyst to use national statistics about industrial production and market shares, to make systematic comparisons among nations (Lundvall, 1992; Nelson, 1993), and to translate the findings into advice for national governments. The relevant statistics have been made comparable among nations by the OECD and Eurostat (OECD/Eurostat, 1997).

The hypothesis of a transition to a 'knowledge-based economy' implies a systems transformation at the structural level across nations. Following this lead, the focus of the efforts at the OECD and Eurostat has been to develop indicators of the relative knowledge-intensity of industrial sectors (OECD, 2001, 2003) and regions (Laafia, 1999, 2002a, 2002b). Alternative frameworks for 'systems of innovation' like technologies (Carlsson & Stankiewicz, 1991) or regions (Braczyk *et al.*, 1998), were also considered (Carlsson, 2004). However, the analysis of the knowledge base of innovation systems (e.g., Cowan *et al.*, 2000) was not made sufficiently relevant for the measurement efforts (David & Foray, 2002). Knowledge was not considered as a coordination mechanism of society, but mainly as a public or private good.

Knowledge as a coordination mechanism was initially defined in terms of the qualifications of the labour force. Machlup (1962) argued that in a 'knowledge economy' knowledge-workers would play an increasingly important role in industrial production processes.

Employment data have been central to the study of this older concept. For example, employment statistics can be cross-tabled with distinctions among sectors in terms of high- and medium-tech (Cooke, 2002; Schwartz, 2006). However, the concept of a 'knowledge-based economy' refers to a change in the structure of an economy beyond the labour market (Foray & Lundvall, 1996; Cooke & Leydesdorff, 2006). How does the development of science and technology transform economic exchange processes? (Schumpeter, 1939).

The social organization of knowledge production and control was first considered as a systemic development by Whitley (1984). Dasgupta & David (1994) proposed to consider science as the subject of a new economics. Because of the reputational control mechanisms involved, the dynamics of knowledge production and diffusion are different in important respects from economic market or institutional control mechanisms (Mirowski & Sent, 2001; Whitley, 2001). When a third coordination mechanism is added as a subdynamic to the interactions and potential co-evolution between economic exchange relations and institutional control (Freeman & Perez, 1988), non-linear effects can be expected (Leydesdorff, 1994). The possible synergies may lead to the envisaged transition to a knowledge-based economy, but this can be expected to happen to a variable extent: developments in some geographically defined economies will be more knowledge-based than others.

The geographical setting, the (knowledge-based) technologies as deployed in different sectors, and the organizational structures of the industries constitute three relatively independent sources of variance. One would expect significant differences in the quality of innovation systems among regions and industrial sectors in terms of technological capacities (Fritsch, 2004). The three sources of variance may reinforce one another in a configuration so that the uncertainty is reduced at the systems level. A knowledge-based order of the economy can thus be shaped.

Our research question is whether one is able to operationalize this configurational order and then also to measure it. For the operationalization we shall need elements from the two research programs which have supported this collaboration: economic geography and scientometrics. We use Storper's (1997, at pp. 26 ff.) notion of a 'holy trinity' among technology, organization, and territory from regional economics, and we elaborate on Leydesdorff's (1995) use of information theory in scientometrics.

## 2. A combination of two theoretical perspectives

Storper (1997, at p. 28) defined a territorial economy as *stocks of relational assets*. The relations determine the dynamics of the system:

> Territorial economies are not only created, in a globalizing world economy, by proximity in input-output relations, but more so by proximity in the untraded or relational dimensions of organizations and technologies. Their principal assets—because scarce and slow to create and imitate—are no longer material, but relational. (Storper, 1997: 28)

The 'holy trinity' is to be understood not only in terms of elements in a network, but as the result of the dynamics of these networks shaping new worlds. These worlds emerge as densities of relations that can be developed into a competitive advantage when and where they materialize by being coupled to the ground in regions. For example, one would expect the clustering of high-tech services in certain (e.g., metropolitan) areas. The location of such a niche can be considered as a consequence of the self-organization of the interactions (Bathelt, 2003; Cooke & Leydesdorff, 2006). Furthermore, Storper argued that this extension of the 'heterodox paradigm' in economics implies a reflexive turn.

In a similar vein, authors using the model of a triple helix of university-industry-government relations have argued for considering the possibility of an overlay of relations among universities, industries, and governments to emerge from these interactions (Etzkowitz & Leydesdorff, 2000). Under certain conditions the feedback from the reflexive overlay can reshape the network relations from which it emerged. Because of this reflexive turn, the parties involved may become increasingly aware of their own and each others' expectations, limitations, and positions. These expectations and interactions can be further informed by relevant knowledge. Thus, the knowledge-based subdynamic may increasingly contribute to the operation of the system.

The triple helix model of university-industry-government relations has hitherto been developed mainly as a *(neo)institutional* model for studying the knowledge infrastructure in networks of relations (Etzkowitz *et al*., 2000; Powell & DiMaggio, 1991). From a *(neo)evolutionary* perspective, a triple helix can be formulated dynamically as the interactions among three (or more) subdynamics of a system (Leydesdorff, 1997; Leydesdorff & Etzkowitz, 1998). To what extent do the networks allow for a synergy among (1) economic wealth generation, (2) technological novelty production, and (3) institutionally organized retention? How can the economic exchange, the innovation dynamics upsetting the market equilibria, and the locally organized interfaces among these subdynamics be integrated at a systems level?

The knowledge-based overlay and the institutional layer operate upon one another in terms of frictions that provide opportunities for innovation both vertically within each of the helices and horizontally among them. The quality of the knowledge base in the economy depends on the locally specific functioning of the interactions in the knowledge infrastructure and on the interface between this infrastructure with the self-organizing dynamics at the systems level. A knowledge base would operate by diminishing the uncertainty that prevails at the network level, that is, as a structural property of the system.

The correspondence between these two perspectives can be extended to the operationalization. Storper (1997, at p. 49), for example, used the following depiction of 'the economy as a set of intertwined, partially overlapping domains of action' in terms of recursively overlapping Venn-diagrams:

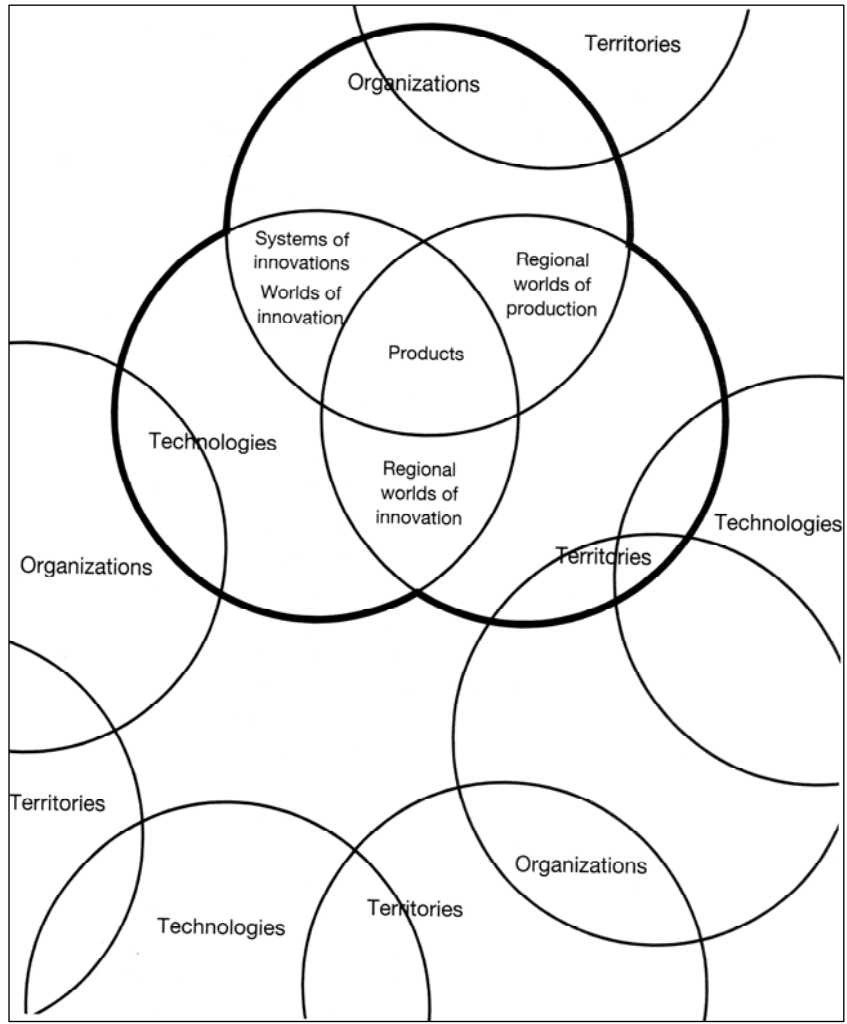

**Figure 1:** Storper's 'holy trinity of technologies, organizations, and territories' provides an overlap in the resulting 'products'.

Using the triple helix model, Leydesdorff (1997, at p. 112) noted that the three circles boldfaced in Figure 1 do not have to overlap in a common zone like the area indicated with 'Products.' He proposed the following configuration as an alternative:

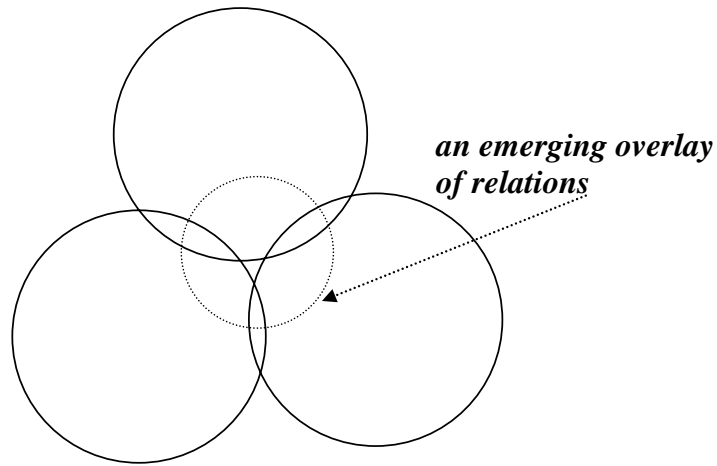

**Figure 2:** The neo-evolutionary variant of the Triple Helix model

In a networked arrangement an overlay of interrelations among the bilateral relations at interfaces can to a certain extent replace the function of central integration. In this case, a virtual hyper-cycle potentially integrates the complex system by providing an additional structure.

The gap in the overlap between the three circles in Figure 2 can be understood as a negative entropy, that is, a reduction of the uncertainty in the system. Unlike the mutual information in two dimensions (Shannon, 1948; Theil, 1972), information among three dimensions thus can become negative (McGill, 1954; Abramson, 1963). This reduction of the uncertainty is in this case a consequence of the networked configuration. However, the 'configurational information' is not present in any of the subsets (Jakulin & Bratko, 2004).[4] In other words, the overlay can be considered as an additional source or sink of information. A configurational reduction of the uncertainty locally counteracts the prevailing tendency at the systems level towards increasing entropy and equilibrium (Khalil, 2004).[5]

---

[4] The so-called interaction or configurational information is defined by these authors as the mutual information in three dimensions, but with the opposite sign (McGill, 1954; Han, 1980).

Using scientometric indicators like co-authorship relations, it could be shown that the configurational information among the three helices varies considerably among nations and world regions (Leydesdorff, 2003). In this study, the scientometric approach to measuring the knowledge base is extended to economic regions at the national and subnational levels by applying the operationalization to data about the distribution of Dutch firms in three dimensions. The data have been used extensively in previous studies by the other authors of this study (Van der Panne & Dolfsma, 2001, 2003; Van der Panne, 2004).

**3. Methods and data**

*3.1    Data*

The data consist of 1,131,668 records containing information based on the registration of enterprises by the Chambers of Commerce of the Netherlands. The data were collected by Marktselect plc on a quarterly basis. Our data specifically correspond to the CD-Rom for the second quarter of 2001 (Van der Panne & Dolfsma, 2003). Because registration with the Chamber of Commerce is obligatory for corporations, the dataset covers the entire population. We brought the data under the control of a relational database manager in order to enable us to focus on the relations more than on the attributes. Dedicated programs were developed for the further processing and computation where necessary.

---

[5] Theil & Fiebig (1984: 12; cf. Frenken, 2000: 263; Sahal, 1979: 129) defined the mutual information in more dimensions as a straightforward extension of the mutual information in two dimensions. However, the decomposition of the mutual information in two and three dimensions enables us to account for the configuration of bilateral and trilateral relations in Triple Helix configurations (Leydesdorff, 2003).

In addition to information at the company level, the data contain three variables which can be used as proxies for the dimensions of technology, organization, and geography at the systems level. Technology will be indicated by the sector classification (Pavitt, 1984; Vonortas, 2000), organization by the company size in terms of numbers of employees (Pugh *et al*., 1969a, 1969b; Blau & Schoenherr, 1971), and the geographical position by the postal codes in the addresses. Sector classifications are based on the European NACE system. This classification was further elaborated by the Dutch Chambers of Commerce into a five-digit system (BIK-codes).[6] In addition to major activities, most companies also provide information about second and third classification terms. However, we shall focus below on the main code using the two-digit level unless otherwise indicated.

The distribution by company size is provided in Table 1. The data contain a first class of 223,231 companies without employees. We decided to include this category because it contains, among others, spin-off companies which are already on the market, but whose owners are employed by mother companies or universities. Given our research question, these establishments are relevant economic activities.

---

[6] NACE stands for Nomenclature générale des Activités économiques dans les Communautés Européennes. The NACE code can be translated into the International Standard Industrial Classificiation (ISIC) and in the Dutch national SBI (Standaard Bedrijfsindeling) developed by Statistics Netherlands. The Chambers of Commerce have elaborated this classification into the so-called BIK code (Bedrijfsindeling Kamers van Koophandel). However, these various codes can be translated unambiguously into one another.

| Size | Number of employees | Number of companies |
|---|---|---|
| 1 | None | 223,231 |
| 2 | 1 | 453,842 |
| 3 | 2 to 4 | 279,835 |
| 4 | 5 to 9 | 88,862 |
| 5 | 10 to 19 | 42,047 |
| 6 | 20-49 | 27,246 |
| 7 | 50-99 | 8,913 |
| 8 | 100-199 | 4,303 |
| 9 | 200-499 | 2,313 |
| 10 | 500-749 | 503 |
| 11 | 750-999 | 225 |
| 12 | > 1000 | 348 |
| | | N = 1,131,668 |

**Table 1**: Distribution of company data by size.

Postal codes are a fine-grained indicator of geographical location. We used the two-digit level which provides us with 90 districts. Using this information the data can be aggregated into provinces (NUTS-2) and so-called COROP regions. The COROP regions correspond with the NUTS-3 level used for the statistics of the OECD and Eurostat.[7] The Netherlands are thus organized in twelve provinces and forty regions, respectively.

*3.2    Knowledge-intensity and high-tech*

The OECD (1986) first defined knowledge-intensity in manufacturing sectors on the basis of R&D intensity. R&D intensity was defined for a given sector as the ratio of R&D expenditure to value added. Later this method was expanded to take account of the technology embodied in purchases of intermediate and capital goods (Hatzichronoglou, 1997). This new measure could also be applied to service sectors which tend to be technology users rather than technology producers. The discussion continues about how best to delineate knowledge-

---

[7] NUTS stands for Nomenclature des Unités Territoriales Statistiques (Nomenclature of Territorial Units for Statistics). COROP is the abbreviation of the Dutch 'Coordinatiecommissie Regionaal Onderzoeksprogramma.'

intensive services (Laafia, 1999, 2002a, 2002b; OECD, 2001, 2003, at p. 140). The classification introduced in the *2001 STI Scoreboard* will be used here (OECD, 2001, at pp. 137 ff.). The relevant NACE categories for high- and medium-tech are as follows:

| High-tech Manufacturing | Knowledge-intensive Sectors (KIS) |
|---|---|
| **30** Manufacturing of office machinery and computers<br>**32** Manufacturing of radio, television and communication equipment and apparatus<br>**33** Manufacturing of medical precision and optical instruments, watches and clocks<br><br>*Medium-high-tech Manufacturing*<br><br>**24** Manufacture of chemicals and chemical products<br>**29** Manufacture of machinery and equipment n.e.c.<br>**31** Manufacture of electrical machinery and apparatus n.e.c.<br>**34** Manufacture of motor vehicles, trailers and semi-trailers<br>**35** Manufacturing of other transport equipment | **61** Water transport<br>**62** Air transport<br>**64** Post and telecommunications<br>**65** Financial intermediation, except insurance and pension funding<br>**66** Insurance and pension funding, except compulsory social security<br>**67** Activities auxiliary to financial intermediation<br>**70** Real estate activities<br>**71** Renting of machinery and equipment without operator and of personal and household goods<br>**72** Computer and related activities<br>**73** Research and development<br>**74** Other business activities<br>**80** Education<br>**85** Health and social work<br>**92** Recreational, cultural and sporting activities<br><br>Of these sectors, **64, 72** and **73** are considered *high-tech services*. |

**Table 2**: Classification of high-tech and knowledge-intensive sectors according to Eurostat. Source: Laafia, 2002a, at p. 7.

These classifications are based on normalizations across the member states of the European Union and the OECD, respectively. However, the percentages of R&D and therefore the knowledge-intensity at the sectoral level may differ in the Netherlands from the average for the OECD or the EU. In a recent report, Statistics Netherlands (CBS, 2003) provided figures for R&D intensity as percentages of value added in 2001. Unfortunately, the data are aggregated at a level higher than the categories provided by Eurostat and the OECD. For this reason, and, furthermore because the Dutch economy is heavily internationalized so that knowledge can easily spill over from neighboring countries, we decided to use the Eurostat categories provided in Table 2 to distinguish levels of knowledge-intensity among sectors.

## 3.3 Regional differences

The reader may need some descriptive statistics to understand the context, since the geographical make-up of the Netherlands is different from its image. The share of employment in high-tech and medium-tech manufacturing in the Netherlands rates only above Luxembourg, Greece, and Portugal in the EU-15 (OECD, 2003, at pp. 140f.). The economically leading provinces of the country, like North- and South-Holland and Utrecht, rank among the lowest on this indicator in the EU-15.[8] The south-east part of the country is integrated in terms of high- and medium-tech manufacturing with neighbouring parts of Belgium and Germany. More than 50% of private R&D in the Netherlands is located in the regions of Southeast North-Brabant and North-Limburg (Wintjes & Cobbenhagen, 2000).

The core of the Dutch economy has traditionally been concentrated on services. These sectors are not necessarily knowledge-intensive, but the situation is somewhat brighter with respect to knowledge-intensive services than in terms of knowledge-based manufacturing. Utrecht and the relatively recently reclaimed province of Flevoland score high on this employment indicator,[9] while both North- and South-Holland are in the middle range. South-Holland is classified as a leading EU region in knowledge-intensive services (in absolute numbers), but the high-tech end of these services has remained underdeveloped. In summary, the country is not homogenous on any of these indicators. On the basis of these employment statistics, the geographical distribution seems almost opposite for high-tech manufacturing and knowledge-intensive services, with provinces specialized in one of the two.

---

[8] Laafia (1999) provides maps of Europe with indication of employment rates in high-tech manufacturing sectors and high-tech service sectors respectively. Laafia (2002a) adds relevant figures.
[9] Flevoland is the only Dutch province amenable for EU support through the structural funds.

*3.4    Methodology*

Unlike a covariation between two variables, a dynamic interaction among three dimensions can generate a complex pattern (Schumpeter, 1939, at pp. 174f; Li & Yorke, 1975). The two configurations possible among three subdynamics were depicted above as integrating or differentiating (in Figures 1 and 2, respectively). In the case of overlapping Venn diagrams, the dynamics can be considered as relatively integrated, e.g., in the resulting products (Storper, 1997, at p. 49; see Figure 1), while in the absence of overlap the system remains more differentiated. In this latter case, it operates in terms of different systems interfacing each other at the network level. In other words, the overlap among the three domains has become negative, and by using the mutual information this can be indicated as negative entropy.[10] Negative entropy reduces the uncertainty that prevails.

Information that is shared among three dimensions can thus be used to measure the extent of integration and differentiation in the interaction among three subsystems. In general, two interacting systems determine each other in their mutual information and condition each other in the remaining uncertainty. They reduce the uncertainty on either side with the mutual information or the transmission. Using Shannon's formulas, this mutual information is defined as the difference between the sum of the uncertainty in two systems without the interaction ($H_x + H_y$) minus the uncertainty contained in the two systems when they are combined ($H_{xy}$). This can be formalized as follows:

$$T_{xy} = H_x + H_y - H_{xy} \qquad (1)$$

---

[10] The relation between the geometrical metaphor of overlap or overlay and the algorithmic measure of mutual information is not strictly one-to-one, but the metaphor is helpful for the understanding.

$H_x$ is the uncertainty in the distribution of the variable $x$ (that is, $H_x = - \sum_x p_x \, {}^2\log p_x$), and analogously, $H_{xy}$ is the uncertainty in the two-dimensional probability distribution (matrix) of $x$ and $y$ (that is, $H_{xy} = - \sum_x \sum_y p_{xy} \, {}^2\log p_{xy}$). The mutual information will be indicated with the T of transmission. If the basis two is used for the logarithm all values are expressed in bits of information.

Abramson (1963, at p. 129) derived from the Shannon formulas that the mutual information in three dimensions is:

$$T_{xyz} = H_x + H_y + H_z - H_{xy} - H_{xz} - H_{yz} + H_{xyz} \qquad (2)$$

While the bilateral relations between the variables reduce the uncertainty, the trilateral integration (represented above as the overlap among the Venn diagrams) adds to the uncertainty. The layers thus alternate in terms of the sign. The sign of $T_{xyz}$ depends on the magnitude of $H_{xyz}$ relative to the mutual information in the bilateral relations.

For example, the trilateral coordination can be associated with a new coordination mechanism that is added to the system. In Figure 1, Storper (1997) indicated the positive overlap with 'Products.' In the network mode (Figure 2), however, a system without strong integration in a center reduces uncertainty by providing a differentiated configuration. The puzzles of integration at the interfaces are then solved in a non-hierarchical, that is, reflexive or knowledge-based mode.

# 4. Results

Let us apply this measure to the data. We will first provide the descriptive statistics (Table 3). As noted, the data allow us to disaggregate in terms of geographical regions (NUTS-2 and NUTS-3), and we are able to distinguish high-tech, medium-tech sectors, and knowledge-intensive services. The various dimensions can also be combined in order to compute the mutual information in a next step (Table 4).

## 4.1 Descriptive statistics

|  | $H_{Geography}$ | $H_{Technology}$ | $H_{Organization}$ | $H_{GT}$ | $H_{GO}$ | $H_{TO}$ | $H_{GTO}$ | N |
|---|---|---|---|---|---|---|---|---|
| NL | 6.205 | 4.055 | 2.198 | 10.189 | 8.385 | 6.013 | 12.094 | 1131668 |
| % $H_{max}$ | *95.6* | *69.2* | *61.3* | *82.5* | *83.2* | *63.7* | *75.9* | |
| Drenthe | 2.465 | 4.134 | 2.225 | 6.569 | 4.684 | 6.039 | 8.413 | 26210 |
| Flevoland | 1.781 | 4.107 | 2.077 | 5.820 | 3.852 | 6.020 | 7.697 | 20955 |
| Friesland | 3.144 | 4.202 | 2.295 | 7.292 | 5.431 | 6.223 | 9.249 | 36409 |
| Gelderland | 3.935 | 4.091 | 2.227 | 7.986 | 6.158 | 6.077 | 9.925 | 131050 |
| Groningen | 2.215 | 4.192 | 2.220 | 6.342 | 4.427 | 6.059 | 8.157 | 30324 |
| Limburg | 2.838 | 4.166 | 2.232 | 6.956 | 5.064 | 6.146 | 8.898 | 67636 |
| N-Brabant | 3.673 | 4.048 | 2.193 | 7.682 | 5.851 | 6.018 | 9.600 | 175916 |
| N-Holland | 3.154 | 3.899 | 2.116 | 6.988 | 5.240 | 5.730 | 8.772 | 223690 |
| Overijssel | 2.747 | 4.086 | 2.259 | 6.793 | 5.002 | 6.081 | 8.749 | 64482 |
| Utrecht | 2.685 | 3.956 | 2.193 | 6.611 | 4.873 | 5.928 | 8.554 | 89009 |
| S-Holland | 3.651 | 3.994 | 2.203 | 7.582 | 5.847 | 5.974 | 9.528 | 241648 |
| Zeeland | 1.802 | 4.178 | 2.106 | 5.941 | 3.868 | 6.049 | 7.735 | 24339 |

**Table 3**: Expected information contents (in bits) of the distributions in the three dimensions and their combinations.

Table 3 shows the probabilistic entropy values in the three dimensions ($G$ = geography, $T$ = technology/sector, and $O$ = organization) for the Netherlands as a whole and the decomposition at the NUTS-2 level of the provinces. The provinces are very different in terms of the numbers of firms and their geographical distribution over the postal codes. While Flevoland contains only 20,955 units, South-Holland provides the location for 241,648 firms.[11] This size effect is reflected in the distribution of postal codes: the uncertainty in the geographical distribution—measured as $H_{Geography}$—correlates significantly with the number of firms N ($r = 0.76$; $p = 0.005$). The variance in the probabilistic entropies among the provinces is high (> 0.5) in this geographical dimension, but the variance in the probabilistic entropy among sectors and the size categories is relatively small (< 0.1). Thus, the provinces are relatively similar in terms of their sector and size distributions,[12] and can thus meaningfully be compared.

The second row of Table 3 informs us that the probabilistic entropy in the postal codes of firms is larger then 95% of the maximum entropy of this distribution at the level of the nation. Since the postal codes are more fine-grained in metropolitan than in rural areas, this indicates that the firm-density is not a major source of variance in relation to the population density. However, the number of postal-code categories varies among the provinces and postal codes are nominal variables which cannot be compared across provinces or regions.

The corresponding percentages for the technology (sector) and the organization (or size) distributions are 69.2 and 61.3%, respectively. The combined uncertainty of technology and organization ($H_{TO}$) does not add substantially to the redundancy. In other words, organization and technology have a relatively independent influence on the distribution different from that of postal codes. In the provincial decomposition, however, the highly developed and densely populated provinces (North and South-Holland, and Utrecht) show a more specialized pattern of sectoral composition ($H_T$) than Friesland, Groningen, and Limburg. (These latter provinces are further distanced from the center of the country.) Flevoland shows the highest redundancy in the size distribution ($H_O$), perhaps because certain traditional formats of middle-sized companies may still be underrepresented in this new province.

---

[11] The standard deviation of this distribution is 80,027.04 with a means of 94,305.7.
[12] The value of H for the country corresponds to the mean of the values for the provinces in these dimensions: $\overline{H}_T = 4.088 \pm 0.097$ and $\overline{H}_O = 2.196 \pm 0.065$.

The combination of technological and organizational specialization exhibits a specific position of North-Holland ($H_{TO}$ = 5.730 or 60.7% of the maximum entropy) versus Friesland ($H_{TO}$ = 6.223 or 65.9% of the maximum entropy) at the other end of the distribution. Since the mean of the distribution is in this case 63.8% with a standard deviation of 1.3, North-Holland is really an exception in terms of an interaction effect between the technological specialization and its relatively low variation in the size distribution.

*4.2   The mutual information*

Table 4 provides the values for the transmissions (T) among the various dimensions. These values can be calculated straightforwardly from the values of the probabilistic entropies provided in Table 3 using Equations 1 and 2 provided above. The first line for the Netherlands as a whole shows that there is more mutual information between the geographical distribution of firms and their technological specialization ($T_{GT}$ = 0.072 bits) than between the geographical distribution and their size ($T_{GO}$ = 0.019). However, the mutual information between technology and organization ($T_{TO}$ = 0.240) is larger than $T_{GO}$ by an order of magnitude. The provinces exhibit a comparable pattern.

|            | $T_{GT}$ | $T_{GO}$ | $T_{TO}$ | $T_{GTO}$ |
|------------|----------|----------|----------|-----------|
| NL         | 0.072    | 0.019    | 0.240    | -0.034    |
| Drenthe    | 0.030    | 0.005    | 0.320    | -0.056    |
| Flevoland  | 0.068    | 0.006    | 0.164    | -0.030    |
| Friesland  | 0.054    | 0.008    | 0.274    | -0.056    |
| Gelderland | 0.040    | 0.004    | 0.242    | -0.043    |
| Groningen  | 0.065    | 0.007    | 0.353    | -0.045    |
| Limburg    | 0.047    | 0.006    | 0.251    | -0.033    |
| N-Brabant  | 0.039    | 0.016    | 0.223    | -0.036    |
| N-Holland  | 0.065    | 0.030    | 0.285    | -0.017    |
| Overijssel | 0.040    | 0.004    | 0.263    | -0.035    |
| Utrecht    | 0.031    | 0.005    | 0.221    | -0.024    |
| S-Holland  | 0.062    | 0.006    | 0.223    | -0.027    |
| Zeeland    | 0.038    | 0.039    | 0.234    | -0.039    |

**Table 4**: The mutual information in two and three dimensions disaggregated at the NUTS 2-level (provinces).

While the values for $T_{GT}$ and $T_{GO}$ can be considered as indicators of the geographical clustering of economic activities (in terms of technologies and organizational formats, respectively), the $T_{TO}$ provides an indicator for the correlation between the maturity of the industry (Anderson & Tushman, 1991) and the specific size of the firms involved (Suárez & Utterback 1995, Utterback & Suárez 1993; cf. Nelson, 1994). The relatively low value of this indicator for Flevoland indicates that the techno-economic structure of this province is less mature than in other provinces. The high values of this indicator for Groningen and Drenthe indicates that the techno-economic structure in these provinces is perhaps relatively over-mature. This indicator can thus be considered as representing a strategic vector (Abernathy & Clark, 1985; Watts & Porter, 2003).

All values for the mutual informations in three dimensions ($T_{TGO}$) are negative. When decomposed at the NUTS-3 level of regions, these values are also negative, with the exception of two regions that contain only a single postal code at the two digits level. (In these two cases the uncertainty is by definition zero.)[13] At first glance, the figures suggest an inverse relationship between the mutual information in three dimensions and the intuitively expected knowledge intensity of regions and provinces, with North-Holland, Utrecht, and South-Holland at the one end and Drenthe and Friesland at the other. However, these values cannot be compared among geographical units without a further normalization. As noted, the postal codes are nominal variables. In a next section, we will focus on the relative effects of decompositions in terms of high- and medium-tech sectors on the geographical units of analysis, but let us first turn to the normalization in the geographical dimension because this dimension provides us with recognizable units (like provinces and regions) which may allow us to validate the indicator.

**5. The regional contributions to the knowledge base of the Dutch economy**

One of the advantages of statistical decomposition analysis is the possibility to specify the within-group variances and the between-group variances in great detail (Theil, 1972; Leydesdorff, 1995). However, a full decomposition at the lower level is possible only if the categories for the measurement are similar among the groups. Had we used a different indicator for the regional dimension—for example, percentage 'rural' versus percentage 'metropolitan'—we would have been able to compare and therefore to decompose along this axis, but the unique postal codes cannot be compared among regions in a way similar to the size or the sectoral distribution of the firms (Leydesdorff & Fritsch, in preparation).

The decomposition algorithm (Theil, 1972) enables us to study the next-order level of the Netherlands as a composed system (NUTS-1) in terms of its lower-level units like the NUTS-

---

[13] These are the regions Delfzijl and Zeeuwsch-Vlaanderen (COROP / NUTS-3 regions 2 and 31).

2 provinces and the NUTS-3 regions. Note that in this case, the regions and provinces are not compared in terms of their knowledge intensity among themselves, but in terms of their weighted contributions to the knowledge base of the Dutch economy as a whole. The distributions are weighted in the various dimensions for the number of firms in the groups $i$ by summing first the uncertainties within the different groups ($\sum_i (n_i/N) * H_i; N = \sum_i n_i$). The in-between group uncertainty $H_0$ is then defined as the difference between this sum and the uncertainty prevailing at the level of the composed system:

$$H = H_0 + \sum_i (n_i/N) H_i \qquad (3)$$

Or equivalently for the transmissions:[14]

$$T = T_0 + \sum_i (n_i/N) T_i \qquad (4)$$

For example, if we use the right-most column of Table 3 indicating the number of firms in each of the provinces for the normalization given the total number of firms registered (N = 1,131,668), we obtain the following table for the decomposition of the mutual information in three dimensions at the level of the provinces:

---

[14] The formula is equally valid for the transmissions because these are based on the probability distributions in the mutual information between two or more probability distributions. The probability distribution in the transmission $T_{ab}$ can be written as the intersect between the distributions for $a$ and $b$, or in formula format as $\sum p_T = \sum (p_a \text{ AND } p_b)$.

|  | $\Delta T_{GTO}$ (= $n_i * T_i /N$) in millibits of information | $n_i$ |
|---|---:|---:|
| Drenthe | -1.29 | 26210 |
| Flevoland | -0.55 | 20955 |
| Friesland | -1.79 | 36409 |
| **Gelderland** | **-4.96** | 131050 |
| Groningen | -1.20 | 30324 |
| Limburg | -1.96 | 67636 |
| **N-Brabant** | **-5.56** | 175916 |
| **N-Holland** | **-3.28** | 223690 |
| Overijssel | -1.98 | 64482 |
| Utrecht | -1.86 | 89009 |
| **S-Holland** | **-5.84** | 241648 |
| Zeeland | -0.83 | 24339 |
| Sum ($\sum_i P_i T_i$) | -31.10 | 1131668 |
| $T_0$ | **-2.46** | |
| NL | -33.55 | N = 1131668 |

**Table 5**: The mutual information in three dimensions statistically decomposed at the NUTS 2-level (provinces) in millibits of information.

The table shows that the knowledge base of the country is concentrated in South-Holland ($\Delta T = -5.84$ mbits), North-Brabant ($-5.56$), and Gelderland ($-4.96$). North-Holland follows with a contribution of $-3.28$ mbits of information. The other provinces contribute to the knowledge base less than the in-between provinces interaction effect at the national level ($T_0 = -2.46$ mbit). Figures 3 and 4 visualize how the knowledge base of the country is geographically organized at the NUTS-2 and the NUTS-3 level, respectively.

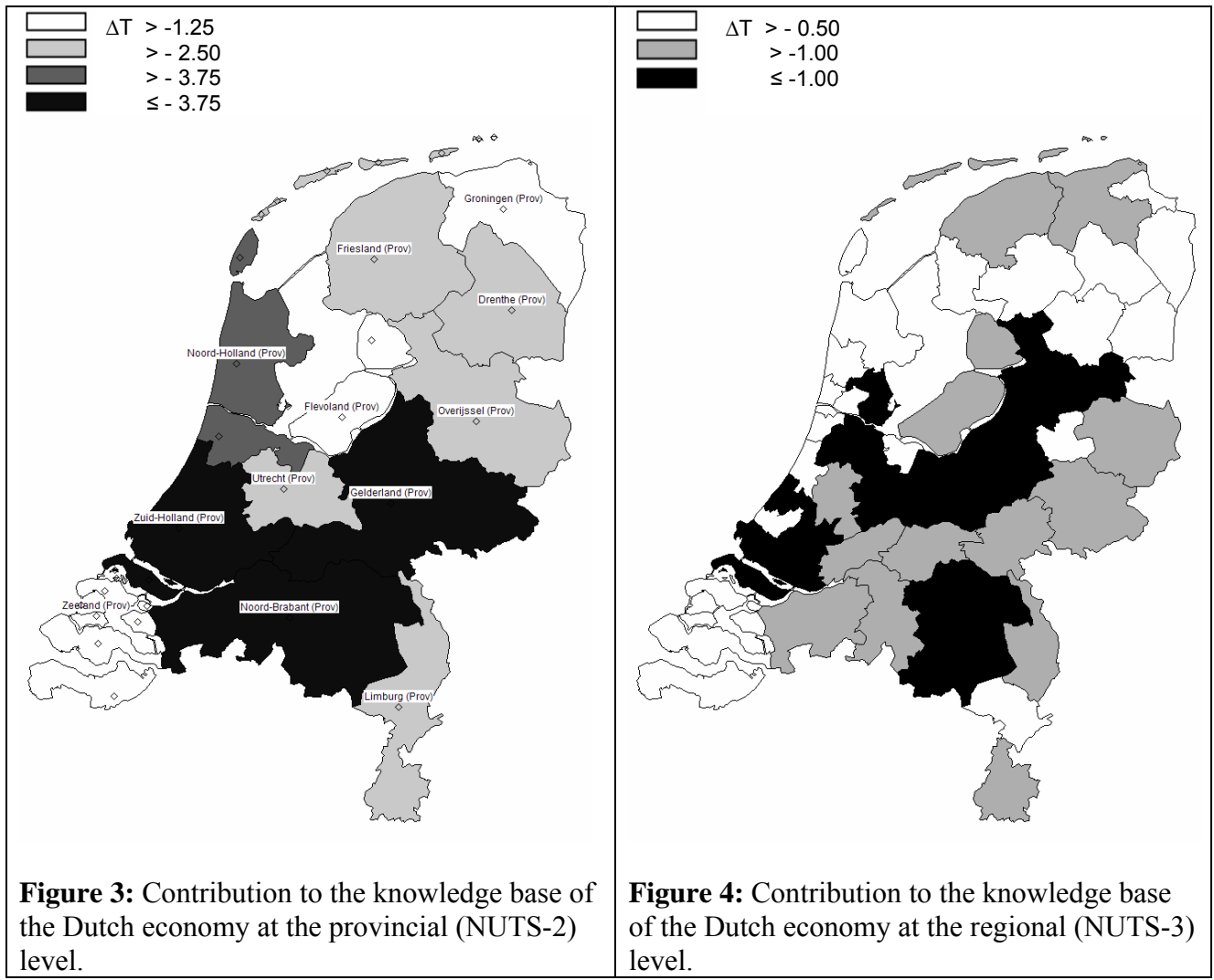

**Figure 3:** Contribution to the knowledge base of the Dutch economy at the provincial (NUTS-2) level.

**Figure 4:** Contribution to the knowledge base of the Dutch economy at the regional (NUTS-3) level.

The further disaggregation in Table 6 informs us about the contribution of regions at the NUTS-3 level (Figure 4). The contribution of South-Holland is concentrated in the Rotterdam area, the one in North-Brabant in the Eindhoven region, and North-Holland exclusively in the agglomeration of Amsterdam. Utrecht, the Veluwe (Gelderland) and the northern part of Overijssel have also above average contributions on this indicator. However, an important part of the reduction of the uncertainty is provided at a level higher than the NUTS-3 regions ($T_0 = -9.09$ mbit).[15] We shall therefore focus in the next section on the NUTS-2 level.

---

[15] More detailed analysis teaches that the provincial structure reduces the uncertainty in the mutual information between the sectoral and the size distribution as two dimensions with $-7.79$ mbits, while this uncertainty is reduced with $-20.06$ mbits by the finer-grained structure of COROP regions. Unlike the effect on the mutual

|    | NUTS-3 Regions (Corop) | $\Delta T_{GTO}$ (= $n_i * T_i / N$) in millibits of information | $n_i$ |
|---|---|---|---|
| 1 | Oost-Groningen | -0.20 | 7571 |
| 2 | Delfzijl en omgeving | 0.00 | 2506 |
| 3 | Overig Groningen | -0.81 | 20273 |
| 4 | Noord-Friesland | -0.99 | 17498 |
| 5 | Zuidwest-Friesland | -0.37 | 7141 |
| 6 | Zuidoost-Friesland | -0.41 | 11744 |
| 7 | Noord-Drenthe | -0.44 | 9702 |
| 8 | Zuidoost-Drenthe | -0.39 | 9121 |
| 9 | Zuidwest-Drenthe | -0.13 | 7327 |
| 10 | **Noord-Overijssel** | **-1.04** | 20236 |
| 11 | Zuidwest-Overijssel | -0.16 | 7333 |
| 12 | Twente | -0.57 | 36971 |
| 13 | **Veluwe** | **-1.38** | 43489 |
| 14 | Achterhoek | -0.76 | 24995 |
| 15 | Arnhem/Nijmegen | -0.85 | 43388 |
| 16 | Zuidwest-Gelderland | -0.69 | 19192 |
| 17 | **Utrecht** | **-1.86** | 88997 |
| 18 | Kop van Noord-Holland | -0.30 | 25978 |
| 19 | Alkmaar en omgeving | -0.39 | 17145 |
| 20 | IJmond | -0.07 | 11017 |
| 21 | Agglomeratie Haarlem | -0.16 | 17376 |
| 22 | Zaanstreek | -0.07 | 9865 |
| 23 | **Groot-Amsterdam** | **-1.15** | 117518 |
| 24 | Het Gooi en Vechtstreek | -0.42 | 24818 |
| 25 | Agglomeratie Leiden en Bollenstreek | -0.42 | 26738 |
| 26 | *Agglomeratie 's-Gravenhage* | *-1.00* | 50603 |
| 27 | Delft en Westland | -0.28 | 19489 |
| 28 | Oost-Zuid-Holland | -0.67 | 25262 |
| 29 | **Groot-Rijnmond** | **-1.61** | 92255 |
| 30 | Zuidoost-Zuid-Holland | -0.91 | 27301 |
| 31 | Zeeuwsch-Vlaanderen | 0.00 | 6840 |
| 32 | Overig Zeeland | -0.39 | 17499 |
| 33 | West-Noord-Brabant | -0.78 | 43954 |
| 34 | Midden-Noord-Brabant | -0.61 | 32332 |
| 35 | *Noordoost-Noord-Brabant* | *-1.00* | 47214 |
| 36 | **Zuidoost-Noord-Brabant** | **-1.13** | 52416 |
| 37 | Noord-Limburg | -0.53 | 16753 |
| 38 | Midden-Limburg | -0.17 | 15272 |
| 39 | Zuid-Limburg | -0.79 | 35611 |
| 40 | Flevoland | -0.55 | 20928 |
|  | Sum ($\sum_i P_i T_i$) | -24.46 | 1131668 |
|  | **$T_0$** | **-9.09** |  |
|  | NL | -33.55 | N = 1131668 |

**Table 6**: The mutual information in three dimensions statistically decomposed at the NUTS-3 level (COROP regions) in millibits of information. Regions with a $\Delta T > 1.00$ mbit are boldfaced; $\Delta T = 1.00$ mbits in italics.

---

information in three dimensions, these reductions of the uncertainty at the NUTS-2 and the NUTS-3 levels are independent of the distribution of postal codes (since specified at these higher levels of aggregation).

These tables and pictures correspond with common knowledge about the industrial structure of the Netherlands (e.g., Van der Panne & Dolfsma, 2001, 2003). The contribution of northern Overijssel to the knowledge base of the Dutch economy is a bit of a surprise because this region has not been recognized previously as an economically active region. Perhaps, it profits from a spill-over effect of knowledge-based activities in the neighbouring regions.

As noted, the normalization involves the number of firms in the geographical unit of analysis as a factor in the weighting. Therefore, these results inform us both about the industrial structure of the country and about the knowledge base of the economy,[16] and may differ depending on the aggregation level analyzed. Among the regions, for example, Utrecht (region 17) contributes most to the reduction of the uncertainty at the national level, while as a province the same value for Utrecht ($\Delta T = -1.86$ mbits) remains below the average contribution. In general, the mutual information in three dimensions provides a composite measure of the three factors involved in Storper's holy trinity (geography, technology, and organization). These three factors can be decomposed along each axis. We turn in the next section to the sectoral axis, and particularly to the effects of indicating knowledge intensity along this axis.

## 6. The sectorial decomposition

While the geographical comparison is compounded with traditional industrial structure like firm density, all effects of the decomposition in terms of the sectorial classification of high- and medium-tech sectors and knowledge-intensive services will be expressed as a relative effect, that is, as a percentage increase or decrease of the negative value of the mutual information in three dimensions when a specific selection is compared with the complete population. In the remainder of this study, we use the categories provided by the OECD and Eurostat (see Table 2 above) as selection criteria for subsets and compare the results with those of the full set provided in the previous section as a baseline. A more negative score for the probabilistic entropy as compared to the overall score indicates a reduction of the uncertainty, and is therefore considered as a more favorable condition for a knowledge-based economy.

---

[16] The correlation between the contributions $\Delta T$ and the number of firms is high and significant both in the case of analysis at the NUTS-2 level ($r = 0.872$; $p < 0.01$) and the NUTS-3 level ($r = 0.801$; $p < 0.01$).

| $T_{xyz}$ | All sectors | High Tech | % change | N |
|---|---|---|---|---|
| NL | -0.034 | -0.060 | 80.2 | 45128 |
| Drenthe | -0.056 | -0.093 | 67.6 | 786 |
| Flevoland | -0.030 | -0.036 | 20.6 | 1307 |
| Friesland | -0.056 | -0.136 | 144.9 | 983 |
| Gelderland | -0.043 | -0.094 | 120.1 | 4885 |
| Groningen | -0.045 | -0.066 | 48.1 | 1204 |
| Limburg | -0.033 | -0.068 | 105.9 | 2191 |
| N-Brabant | -0.036 | -0.058 | 61.2 | 6375 |
| N-Holland | -0.017 | -0.034 | 103.4 | 9346 |
| Overijssel | -0.035 | -0.079 | 127.6 | 2262 |
| Utrecht | -0.024 | -0.039 | 65.9 | 4843 |
| S-Holland | -0.027 | -0.044 | 61.7 | 10392 |
| Zeeland | -0.039 | -0.067 | 73.3 | 554 |

**Table 7**: The mutual information in three dimensions when comparing high-tech sectors in industrial production and services.

Table 7 provides the results of comparing the subset of enterprises indicated as high-tech manufacturing (sectors 30, 32, and 33) and high-tech services (64, 72, and 73) with the full set. The column headed with 'All sectors' corresponds to the right-most column in Table 3. The third column provides the mutual information in three dimensions for the high-tech sectors in both manufacturing and services. In the fourth column the percentage change is indicated in relative terms. This indicates the influence of these high-tech sectors and services on the knowledge base of the economy. The results confirm our hypothesis that the mutual information or entropy that emerges from the interaction between the three dimensions is more negative for high-tech sectors and high-tech services than for the economy as a whole. The dynamics created by these sectors deepen and tighten the knowledge base more than is the case for firms on the average.

|  | All sectors | High & medium tech Manufacturing | % change | N | Knowledge-Intensive Services | % change | N |
|---|---|---|---|---|---|---|---|
| NL | -0.034 | -0.219 | 553 | 15838 | -0.024 | -27.3 | 581196 |
| Drenthe | -0.056 | -0.349 | 526 | 406 | -0.034 | -39.1 | 11312 |
| Flevoland | -0.030 | -0.206 | 594 | 401 | -0.018 | -37.9 | 10730 |
| Friesland | -0.056 | -0.182 | 227 | 951 | -0.037 | -32.6 | 14947 |
| Gelderland | -0.043 | -0.272 | 536 | 2096 | -0.025 | -40.8 | 65112 |
| Groningen | -0.045 | -0.258 | 479 | 537 | -0.029 | -34.0 | 14127 |
| Limburg | -0.033 | -0.245 | 647 | 1031 | -0.018 | -45.1 | 30040 |
| N-Brabant | -0.036 | -0.190 | 430 | 2820 | -0.030 | -16.6 | 86262 |
| N-Holland | -0.017 | -0.173 | 943 | 2299 | -0.017 | 1.0 | 126516 |
| Overijssel | -0.035 | -0.207 | 496 | 1167 | -0.020 | -42.8 | 30104 |
| Utrecht | -0.024 | -0.227 | 859 | 1020 | -0.013 | -45.0 | 52818 |
| S-Holland | -0.027 | -0.201 | 635 | 2768 | -0.015 | -45.5 | 128725 |
| Zeeland | -0.039 | -0.180 | 365 | 342 | -0.028 | -27.8 | 10503 |

**Table 8**: High-tech & medium-tech manufacturing versus knowledge-intensive services and the effects on the mutual information in three dimensions.

Table 8 provides the same figures and normalizations, but now on the basis of selections according to the classifications provided in Table 2 for high- and medium-tech manufacturing combined (middle section of Table 2), and knowledge-intensive services (right-side columns of Table 2), respectively. These results indicate a major effect on the indicator for the sectors of high- and medium-tech manufacturing. The effect is by far the largest in North-Holland with 943% increase relative to the benchmark of all sectors combined. Utrecht follows with 859 %. A group of provinces (Limburg, South-Holland, Flevoland) has above average effects of 647, 635, and 594%, respectively. Zeeland has the lowest value on this indicator (365%), but the number of establishments in these categories is also lowest for this province. North-Brabant, however, has the largest number of establishments in these categories, while it does not seem to profit from an additional effect on the configuration.

The number of establishments in knowledge-intensive services is more than half (51.3%) of the total number of companies in the country. These companies are concentrated in North- and South-Holland, with North-Brabant in the third position. With the exception of North-Holland, the effect of knowledge-intensive services on this indicator of the knowledge base is always negative, that is, it leads to a decrease of configurational information. We indicate this with an opposite sign for the change. In the case of North-Holland, the change is marginally positive (+1.0 %), but this is not due to the Amsterdam region.[17] North-Brabant is second on this rank order with a decrease of −16.6%.

These findings accord with a theoretical expectation about the different contributions to the economy of services in general and KIS in particular (Bilderbeek *et al*., 1998; Miles *et al*., 1995; OECD, 2000). Windrum & Tomlinson (1999) argued that to assess the role of KIS, the degree of integration is more important than the percentage of representation in the economy. Unlike output indicators, the measure adopted here focuses on the degree of integration in the configuration. However, our results indicate that KIS unfavorably affects the synergy between technology, organization, and territory in the techno-economic system of the Netherlands, its provinces, and regions. This indicates a relatively uncoupling effect from the geographically defined knowledge bases of the economy. The effects of KIS are spilling over geographic boundaries more easily than knowledge-based manufacturing.

---

[17] Only in COROP / NUTS-3 region 18 (North-Holland North) the value of the mutual information in three dimensions is more negative when zooming in on the knowledge-intensive services. However, this region is predominantly rural.

This result contrasts with the expectations expressed in much of the relevant literature on the role of knowledge-intensive services in stimulating the knowledge base of an economy. For example, the conclusion of the European Summit in Lisbon (2000) was, among other things, that "the shift to a digital, knowledge-based economy, prompted by new goods and services, will be a powerful engine for growth, competitiveness and jobs. In addition, it will be capable of improving citizens' quality of life and the environment."[18] Our results suggest that the knowledge-based economy and the digital economy are not the same: the manufacturing of goods or the delivering of services can be expected to have other geographical effects and constraints.

Knowledge-intensive services seem to be largely uncoupled from the knowledge flow within a regional or local economy. They contribute negatively to the knowledge-based configuration because of their inherent capacity to deliver these services outside the region. Thus, a locality can be chosen on the basis of considerations other than those relevant for the generation of a knowledge-based economy in the region. For example, the proximity of a well-connected airport (or train station) may be a major factor in the choice of a location.

---

[18] at http://www.europarl.eu.int/summits/lis1_en.htm#b .

| $T_{xyz}$ | Knowl-intensive services | High-Tech services | % change | N |
|---|---|---|---|---|
| NL | -0.024 | -0.034 | 37.3 | 41002 |
| Drenthe | -0.034 | -0.049 | 45.2 | 678 |
| Flevoland | -0.018 | -0.018 | -4.6 | 1216 |
| Friesland | -0.037 | -0.087 | 131.5 | 850 |
| Gelderland | -0.025 | -0.046 | 82.3 | 4380 |
| Groningen | -0.029 | -0.044 | 49.5 | 1070 |
| Limburg | -0.018 | -0.039 | 118.7 | 1895 |
| N-Brabant | -0.030 | -0.035 | 16.1 | 5641 |
| N-Holland | -0.017 | -0.020 | 17.0 | 8676 |
| Overijssel | -0.020 | -0.046 | 133.1 | 1999 |
| Utrecht | -0.013 | -0.020 | 49.8 | 4464 |
| S-Holland | -0.015 | -0.025 | 69.8 | 9650 |
| Zeeland | -0.028 | -0.045 | 59.7 | 483 |

**Table 9**: The subset of high-tech services improves the knowledge base in the service sector.

Table 9 shows the relative deepening of the mutual information in three dimensions when the subset of sectors indicated as 'high-tech services' is compared with KIS in general. 'High-tech services' are only 'post and telecommunications' (NACE code 64), 'computer and related activities' (72), and 'research and development' (73). More than knowledge-intensive services in general, high-tech services can be expected to produce and transfer technology-related knowledge (Bilderbeek *et al.*, 1998). These effects of strengthening the knowledge base seem highest in regions which do not have a strong knowledge base in medium and high-tech manufacturing, such as Friesland and Overijssel. The effects of this selection for North-Brabant and North-Holland, for example, are among the lowest. However, this negative relation between high- and medium-tech manufacturing on the one hand, and high-tech services on the other, is not significant ($r = -0.352$; p = 0.262). At the NUTS-3 level, the corresponding relation is also not significant. Thus, the effects of high- and medium-tech manufacturing and high-tech services on the knowledge base of the economy are not related to each other.

## 7. Conclusions and discussion

Before we proceed to draw conclusions and consider policy implications, we should emphasize that this effort was initially motivated by methodological considerations. We had developed independently, on the one hand, an indicator of interaction effects at the network level which provided us with a quantitative measure for the reduction of the uncertainty that cannot be attributed to the individual players in a network. The reduction of this uncertainty is configurational. It indicates that a next-order system is operating as an overlay. On the other hand, the data and insights from economic geography allowed us to use proxies for the three main dimensions of Storper's 'holy trinity' of technology, organization, and territory, although the operationalization of organization in terms of numbers of employees remains debatable.

Our collaboration provided us with an opportunity to validate the scientometric indicator of triple-helix relations in an economic context. We submit that this indicator can be used to measure the knowledge base of an economy and its decomposition in terms of geographical subunits and in considerable detail. However, the nominal character of the postal codes made it impossible to compare directly among regions and provinces (but only in terms of their contribution to the national economy). The effects of these problems on the data in relation to the indicator and its interpretation made us hesitant to decompose below the NUTS-2 level of the provinces because this may suggest an exactness which cannot be achieved using the data. Nevertheless, our results allow us to formulate the following hypotheses:

1. The knowledge base of a (regional) economy is carried by high-, but more importantly by medium-tech manufacturing; high-tech services favorably contribute to the knowledge-based structuring, but to a smaller extent.
2. Medium-tech manufacturing provides the backbone of the techno-economic structure of the country; this explains why high-tech manufacturing contributes less to the knowledge infrastructure than might be expected, for example, on the basis of patent portfolios (Leydesdorff, 2004).
3. The knowledge-intensive services which are not high-tech have a relatively unfavorable effect on the territorial knowledge base of an economy. One could say that these services tend to uncouple the knowledge base from its geographical dimension.
4. The Netherlands is highly developed as a knowledge-intensive service economy, but the high-tech end of these services has remained more than an order of magnitude smaller in terms of the numbers of firms.

These conclusions can be tested, for example, by using comparable data for another country (Leydesdorff & Fritsch, in preparation). In terms of policy implications, these conclusions suggest that regions which are less developed may wish to strengthen their knowledge infrastructure by trying to attract medium-tech manufacturing and high-tech services. The efforts of firms in medium-tech sectors can be considered as focused on maintaining absorptive capacity (Cohen & Levinthal, 1989) so that knowledge and technologies developed elsewhere can more easily be understood and adapted to particular circumstances. High-tech manufacturing may be more focused on the (internal) production and global markets than on the local diffusion parameters. High-tech services, however, mediate technological knowledge more than knowledge-intensive services which are medium-tech.

The latter services seem to have an unfavorable effect on territorially defined knowledge-based economies.

It should be noted that the indicator measures a synergy at the structural level of an economy and is not a measure of knowledge creation or economic output (Carter, 1996). The synergy among the industrial structures, geographical distributions, and technological capacities can be considered crucial for the strength of an innovation system (Fritsch, 2004). In other words, this indicator measures only the conditions in the system for innovative activities, and thus specifies an expectation (Dolfsma, 2005). Regions with a high potential for innovative activity can be expected to organize more innovative resources than regions with lower values of the indicator.

Perhaps the most important contribution of this paper is the procedure presented for measuring the synergy as the expected knowledge base of an economy. The various dimensions correspond to the classifications that are already available from the OECD and Eurostat databases, and the geographical address information of the units is also used. In principle, the dimensionality of the mutual information can further be extended. Unlike the focus on comparative statics in employment statistics and the *STI Scoreboards* of the OECD (OECD, 2001, 2003; Godin, 2006), this indicator was developed for measuring the knowledge base of an economy as an emergent property (Jakulin & Bratko, 2004; cf. Ulanowicz, 1986, at pp. 142 ff.). Furthermore, the indicator could be specified as an operationalization with reference to two bodies of theorizing in evolutionary economics, namely regional studies (e.g., Storper, 1997; Van der Panne & Dolfsma, 2003) and the study of knowledge-based systems of innovation (e.g., David & Foray, 2002; Leydesdorff & Etzkowitz, 1998).